\documentclass[prl,showpacs,floatfix,twocolumn]{revtex4-1}
\usepackage{amsfonts}
\usepackage{stmaryrd}
\usepackage{bbm}
\usepackage{mathrsfs}
\usepackage{tipa}
\usepackage{amssymb}
\usepackage{txfonts}
\usepackage{graphicx}
\usepackage{dcolumn}
\usepackage{epstopdf}
\usepackage[colorlinks,linkcolor=blue,urlcolor=blue,citecolor=blue]{hyperref}
\newcommand{\PreserveBackslash}[1]{\let\temp=\\#1\let\\=\temp}
\newcolumntype{C}[1]{>{\PreserveBackslash\centering}p{#1}}
\newcolumntype{R}[1]{>{\PreserveBackslash\raggedleft}p{#1}}
\newcolumntype{L}[1]{>{\PreserveBackslash\raggedright}p{#1}}

\begin{document}

\newcommand*{\cm}{cm$^{-1}$\,}

\title{Structural phase transition induced by van Hove singularity in 5\emph{d} transition metal compound IrTe$_{2}$}

\author{T. Qian$^{1}$}
\author{H. Miao$^{1}$}
\author{Z. J. Wang$^{1}$}
\author{X. Liu$^{1,2}$}
\author{X. Shi$^{1}$}
\author{Y. B. Huang$^{1}$}
\author{P. Zhang$^{1}$}
\author{N. Xu$^{1}$}
\author{P. Richard$^{1,3}$}
\author{M. Shi$^{4}$}
\author{M. H. Upton$^{5}$}
\author{J. P. Hill$^{2}$}
\author{G. Xu$^{1}$}
\author{X. Dai$^{1,3}$}
\author{Z. Fang$^{1,3}$}
\author{H. C. Lei$^{2}$}
\author{C. Petrovic$^{2}$}
\author{A. F. Fang$^{1}$}
\author{N. L. Wang$^{1,3}$}
\author{H. Ding$^{1,3}$}


\affiliation{$^1$Beijing National Laboratory for Condensed Matter Physics, and Institute of Physics, Chinese Academy of Sciences, Beijing 100190, China}

\affiliation{$^2$Condensed Matter Physics and Materials Science Department, Brookhaven National Laboratory, Upton, New York 11973, USA}

\affiliation{$^3$Collaborative Innovation Center of Quantum Matter, Beijing, China}

\affiliation{$^4$Paul Scherrer Institute, Swiss Light Source, CH-5232 Villigen PSI, Switzerland}

\affiliation{$^5$Advanced Photon Source, Argonne National Laboratory, Argonne, Illinois 60439, USA}


\begin{abstract}

Comprehensive studies of the electronic states of Ir 5\emph{d} and Te 5\emph{p} have been performed to elucidate the origin of the structural phase transition in IrTe$_{2}$ by combining angle-resolved photoemission spectroscopy and resonant inelastic X-ray scattering. While no considerable changes are observed in the configuration of the Ir 5\emph{d} electronic states across the transition, indicating that the Ir 5\emph{d} orbitals are not involved in the transition, we reveal a van Hove singularity at the Fermi level (\emph{E}$_{F}$) related to the Te \emph{p}$_{x}$+\emph{p}$_{y}$ orbitals, which is removed from \emph{E}$_{F}$ at low temperatures. The wavevector connecting the adjacent saddle points is consistent with the in-plane projection of the superstructure modulation wavevector. These results can be qualitatively understood with the Rice-Scott ``saddle-point'' mechanism, while effects of the lattice distortions need to be additionally involved.

\end{abstract}


\maketitle

The proximity of superconductivity to another quantum ordered state is a hot issue in condensed matter physics. The interplay between them is closely related to the origin of superconductivity. For example, in cuprates, ferropnictides and ferrochalcogenides, and heavy fermion superconductors, the development of superconductivity is usually accompanied by a decline of long-range magnetic order. It is thus widely believed that the origin of their unconventional superconductivity is intimately associated with the magnetic excitations, though the microscopic mechanisms are still a mystery \cite{cuprate,iron,HF}. Another typical example is the coexistence and competition between charge-density-wave (CDW) and superconductivity in a number of compounds, which has attracted much attention to their relationship \cite{TiSe2,NbSe2_Shin,NbSe2_Borisenko,TiS2_FengDL}. 

Recently, the 5\emph{d} transition metal dichalcogenide IrTe$_{2}$ with 1T-structure was suggested to be of the second case. IrTe$_{2}$ exhibits a structural phase transition from trigonal \emph{P}$\overline{3}$\emph{m}1 to triclinic \emph{P}1 at \emph{T}$_{s}$ $\sim$ 270 K \cite{structure}, below which a new structural modulation with a wavevector \emph{Q} = (1/5, 0, -1/5) was revealed \cite{ED1}. With intercalation or substitution at the Ir sites of Pt, Pd, Rh or Cu, the structural transition is suppressed and bulk superconductivity is induced with \emph{T}$_{c}$ up to $\sim$ 3 K, indicating interesting interplay between them \cite{ED1,optical,JPSJ_PS,CuxIrTe2,RhxIrTe2}. Clarifying the origin of the structural transition is a crucial step to understand the mechanism of the superconductivity. Yang \emph{et} \emph{al}. proposed that the transition is driven by partial Fermi surface (FS) nesting, and therefore of the CDW-type with involvement of the Ir 5\emph{d} orbitials \cite{ED1}. However, no energy gap, characteristic of a density-wave type transition, was identified from the optical conductivity spectra \cite{optical}. Several other mechanisms, including the orbital-induced Peierls instability \cite{ARPES,photoemission}, the reduction of kinetic energy by the crystal field effects \cite{optical}, the interlayer hybridization \cite{CuxIrTe2}, the local bonding instability \cite{structure}, and the anionic depolymerization transition \cite{ED2}, have been proposed to understand the structural transition. 

In this Letter, we provide direct evidence that the phase transition in IrTe$_{2}$ is intimately associated with the van Hove singularity (vHs) at the Fermi level (\emph{E}$_{F}$) related to the Te \emph{p}$_{x}$+\emph{p}$_{y}$ orbitals. The vHs arises from six saddle points at \emph{k}$_{z}$ = $\pi$ and the wavevector between the adjacent saddle points is \emph{q} $\sim$ (0.19, 0, 0), which is very close to the in-plane projection of the structural modulation wavevector. The band structure related to the saddle points is dramatically reconstructed below \emph{T}$_{s}$, leading to a significant reduction of the kinetic energy, which is likely to be the driving force for the transition. The results can be qualitatively understood in the framework of the Rice-Scott ``saddle-point'' mechanism.

High quality single crystals of IrTe$_{2}$ were grown via the self-flux technique \cite{optical}. Resonant inelastic X-ray scattering (RIXS) measurements were carried out at beamline 30-ID, Advanced Photon Source, in a horizontal scattering geometry. A Si(844) secondary monochromator and a \emph{R} = 2\emph{m} Si(844) diced analyzer were utilized. The overall energy resolution of this setup was $\sim$ 40 meV (FWHM). Angle-resolved photoemission spectroscopy (ARPES) experiments were performed at beamline PGM of the Synchrotron Radiation Center (Wisconsin), at beamline 4.0.3.2 of the Advanced Light Source (California), and beamline SIS of the Swiss Light Source (Switzerland) with Scienta R4000 analyzers. The energy and angular resolutions were set at 30 meV and 0.2$^{\circ}$, respectively. The samples were cleaved \emph{in} \emph{situ} and measured in a vacuum better than 3$\times$10$^{-11}$ Torr. Calculations for the electronic structure and density of states (DOS) were performed by using the full-potential augmented plane-wave and Perdew-Burke-Ernzerhof parametrization of the generalized gradient approximation (GGA-PBE) exchange-correlation function \cite{GGA} as implemented in the WIEN2k code \cite{WIEN2k}. The spin-orbital interactions were included by using a second variational procedure. The muffin-tin radii (\emph{R}$_{MT}$) were set to 2.50 bohrs for Ir and 2.26 bohrs for Te. The plane-wave cutoff (\emph{K}$_{max}$) was determined by \emph{R}$_{min}$\emph{K}$_{max}$ = 7.0, where \emph{R}$_{min}$ is the minimal \emph{R}$_{MT}$.

\begin{figure}
\includegraphics[clip,width=3in]{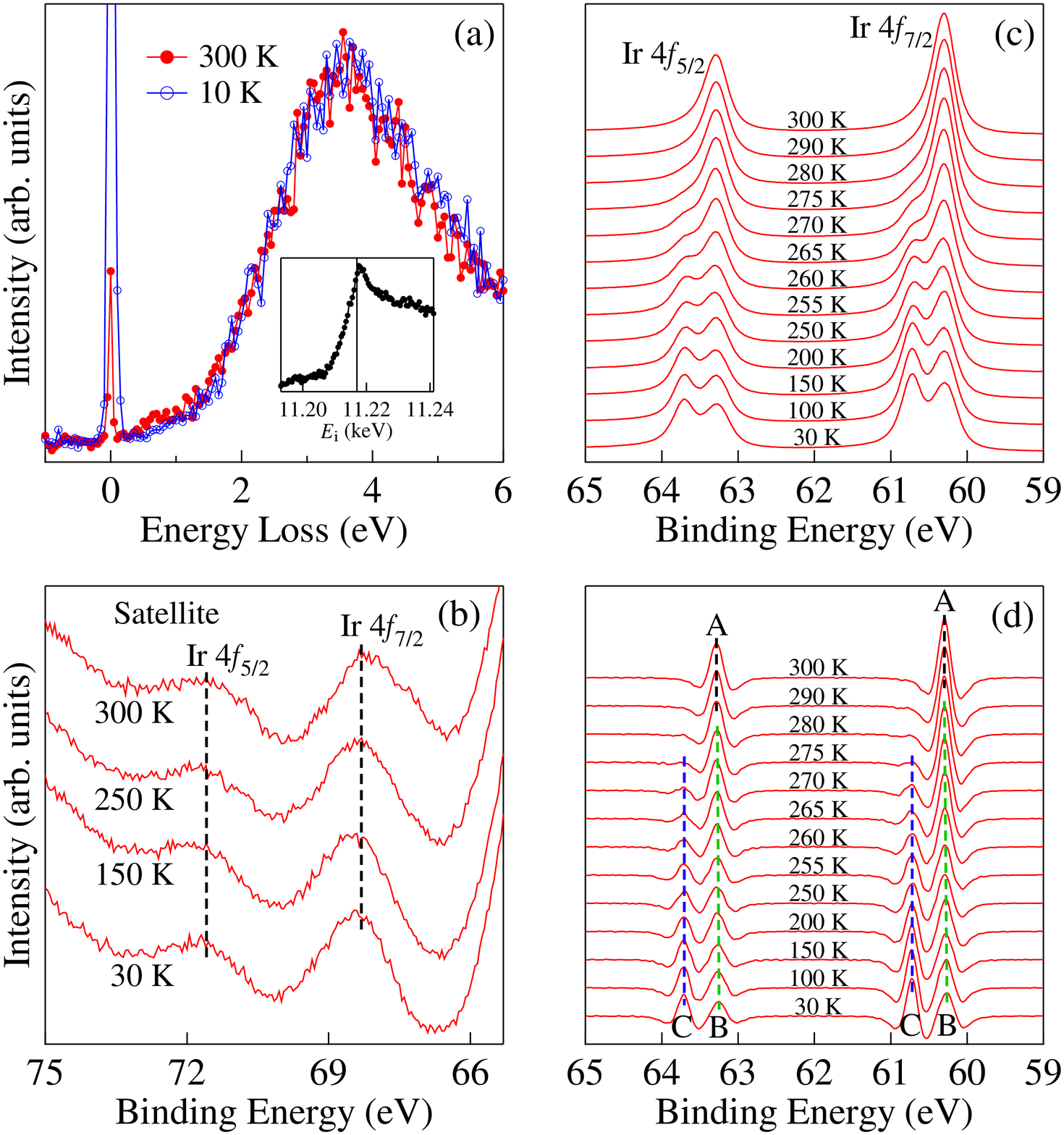}
\caption{(Color online) (a) Ir L3-edge RIXS spectra taken at 300 and 10 K. The inset shows the absorption spectra at the Ir L3-edge with the solid vertical line indicating the energy of the incident X-ray. (b) and (c) Photoemission spectra of the Ir 4\emph{f} satellites and main lines taken with \emph{h}$\nu$ = 140 eV at various temperatures, respectively. (d) Second derivatives of the spectra shown in panel (c). Dashed lines are guides for eyes to trace the peak positions.}
\end{figure}

First we used RIXS and core level photoemission spectroscopy to show that the FSs of IrTe$_{2}$ are dominated by the Te 5\emph{p} orbitals and the phase transition is mainly related to the Te 5\emph{p} electronic states near \emph{E}$_{F}$. To probe the electronic configuration of the Ir 5\emph{d} orbitals, we have carried out RIXS measurements at the Ir-L3 edge. As a function of incident X-ray energy (\emph{E}$_{i}$), a strong resonant peak is found at an energy loss of $\sim$ 3.5 eV [Fig. 1(a)], whose intensity is maximized at \emph{E}$_{i}$ $\sim$ 11.217 keV [inset of Fig. 1(a)]. The peak corresponds to the Ir \emph{t}$_{2g}$ $\rightarrow$ \emph{e}$_{g}$ excitations. If the Ir \emph{t}$_{2g}$ orbitals are partially occupied, excitations within \emph{t}$_{2g}$ orbitals are expected, which resonate at slightly lower \emph{E}$_{i}$. Features related to such excitations are observed in CuIr$_{2}$S$_{4}$, whose temperature dependence has been suggested to be the signature of Ir 3+:4+ charge disproportion \cite{RIXS}. For IrTe$_{2}$, the inelastic response down to 0.2 eV follows the same resonant behavior as the 3.5 eV peak, and shows no indication of excitations within the \emph{t}$_{2g}$ orbitals. This suggests strongly that the \emph{t}$_{2g}$ orbitals are fully occupied. With the Ir \emph{e}$_{g}$ levels much higher in energy due to the crystal field splitting, the FSs are expected to be dominated by the Te 5\emph{p} orbitals. We compare in Fig. 1(a) the RIXS response above and below \emph{T}$_{s}$ with data collected on the superstructure wavevector (1.6, 0, 6.4). Although the structural transition is clearly shown by the drastic increase of elastic peak at this superstructure wavevector, the inelastic response shows no appreciable difference at all, indicating that the Ir 5\emph{d} orbitals are not involved in the transition.

The above conclusion is further supported by our core-level photoemission spectroscopy measurements. While a previous study has shown an increase in the Ir 4\emph{f} peak width \cite{photoemission}, our high-resolution measurements reveal clear splitting of the Ir 4\emph{f} and Te 4\emph{d} peaks below the phase transition. Figures 1(c) and 1(d) show the main peaks in the Ir 4\emph{f} core level spectra and their second derivatives at various temperatures, respectively. A set of peaks (labeled as \emph{A} in Fig. 1(d)) induced by the spin-orbital coupling are observed above \emph{T}$_{s}$. The binding energy (\emph{E}$_{B}$) of Ir 4\emph{f}$_{7/2}$ in IrTe$_{2}$ is 60.3 eV at 300 K, which is obviously lower than in IrO$_{2}$ (62.0 eV, Ir$^{4+}$) \cite{IrO2} and CuIr$_{2}$S$_{4}$ ($\sim$ 61.0 eV, Ir$^{3.5+}$) \cite{CuIr2S4}. This suggests that the valence of Ir in IrTe$_{2}$ is near 3+, in agreement with the RIXS results. Upon cooling below \emph{T}$_{s}$, they are split into two sets of peaks (\emph{B} and \emph{C}), whose energy difference $\Delta$$_{BC}$ slightly depends on temperature and is $\sim$ 0.44 eV at 30 K. In contrast, the Ir 4\emph{f} satellite peaks in Fig. 1(b) exhibit no considerable change. It is known that the satellite and main peaks correspond to the unscreened 5\emph{d}$^{6}$ and well-screened 5\emph{d}$^{7}$$\underline{\emph{L}}$ ($\underline{\emph{L}}$ denotes one hole in the ligand Te 5\emph{p} orbitals) final states, respectively. Thus the satellite line energy does not depend on the Te 5\emph{p} electronic states, while the main line energy strongly correlates with them. Since the satellites exhibit no obvious difference, the electronic configuration of the Ir 5\emph{d} orbitals does not change below \emph{T}$_{s}$, in agreement with the RIXS results. Therefore, the splitting in the Ir 4\emph{f} main peaks is due to a reconstruction of the Te 5\emph{p} electronic states.

\begin{figure}
\includegraphics[clip,width=3in]{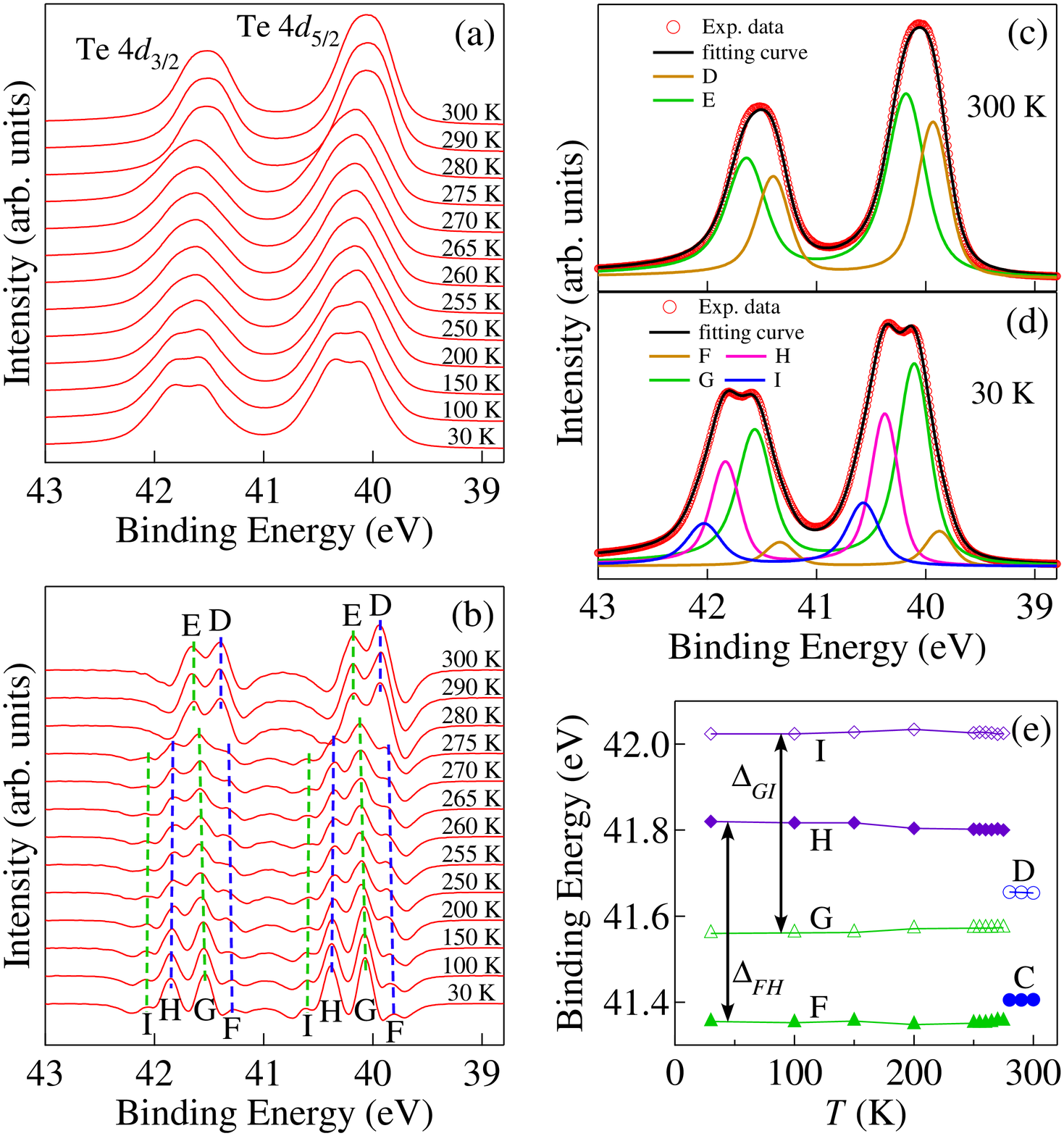}
\caption{(Color online) (a) Photoemission spectra of the Te 4\emph{d} core lines taken with \emph{h}$\nu$ = 90 eV at various temperatures. (b) Second derivatives of the spectra shown in panel (a). Dashed lines are guides for eyes to trace the peak positions. (c) and (d) Te 4\emph{d} photoemission spectra taken at 300 and 30 K with fitted curves, respectively. The spectra at 300 and 30 K are fitted to two and four sets of peaks with the Doniach-Sunjic line shape convoluted by a Gaussian with FWHM of 0.26 eV, respectively. (e) Energy positions of maxima of the \emph{D}, \emph{E}, \emph{F}, \emph{G}, \emph{H} and \emph{I} peaks for the Te 4\emph{d}$_{3/2}$ states, determined from the fitted curves as exemplified in panels (c) and (d), as a function of temperature. $\Delta$$_{FH}$ ($\Delta$$_{GI}$) represents the energy difference between the peaks \emph{F} and \emph{H} (\emph{G} and \emph{I}).}
\end{figure}

The reconstruction of the Te 5\emph{p} electronic states below \emph{T}$_{s}$ is also directly reflected in the Te 4\emph{d} core level spectra, as shown in Fig. 2, where a drastic change is seen across \emph{T}$_{s}$. Two sets of peaks of Te 4\emph{d} (\emph{D} and \emph{E}) are identified above \emph{T}$_{s}$. A possible reason for the double-peak feature is that they may originate from two different screening channels, where the screening charge comes from the Te 5\emph{p} conduction bands and the ligand Ir 5\emph{d} orbitals, respectively. Similar double-peak features have been also observed in La$_{1-x}$Sr$_{x}$MnO$_{3}$ and V$_{2}$O$_{3}$, and interpreted in the same way \cite{LSMO,V2O3}. Upon cooling below \emph{T}$_{s}$, four sets of peaks (\emph{F}, \emph{G}, \emph{H}, and \emph{I}) are identified. The spectra above and below \emph{T}$_{s}$ are fitted to two and four sets of peaks with the Doniach-Sunjic line shape, respectively. The peak positions determined by fitting are plotted as a function of temperature in Fig. 2(e). The energy differences $\Delta$$_{FH}$ and $\Delta$$_{GI}$ are equal in the whole temperature range below \emph{T}$_{s}$ and $\sim$ 0.46 eV at 30 K. The value is very close to that of $\Delta$$_{BC}$ of Ir 4\emph{f} at 30 K, indicating that the splitting in the Ir 4\emph{f} and Te 4\emph{d} peaks has the same origin. 

\begin{figure}
\includegraphics[clip,width=3in]{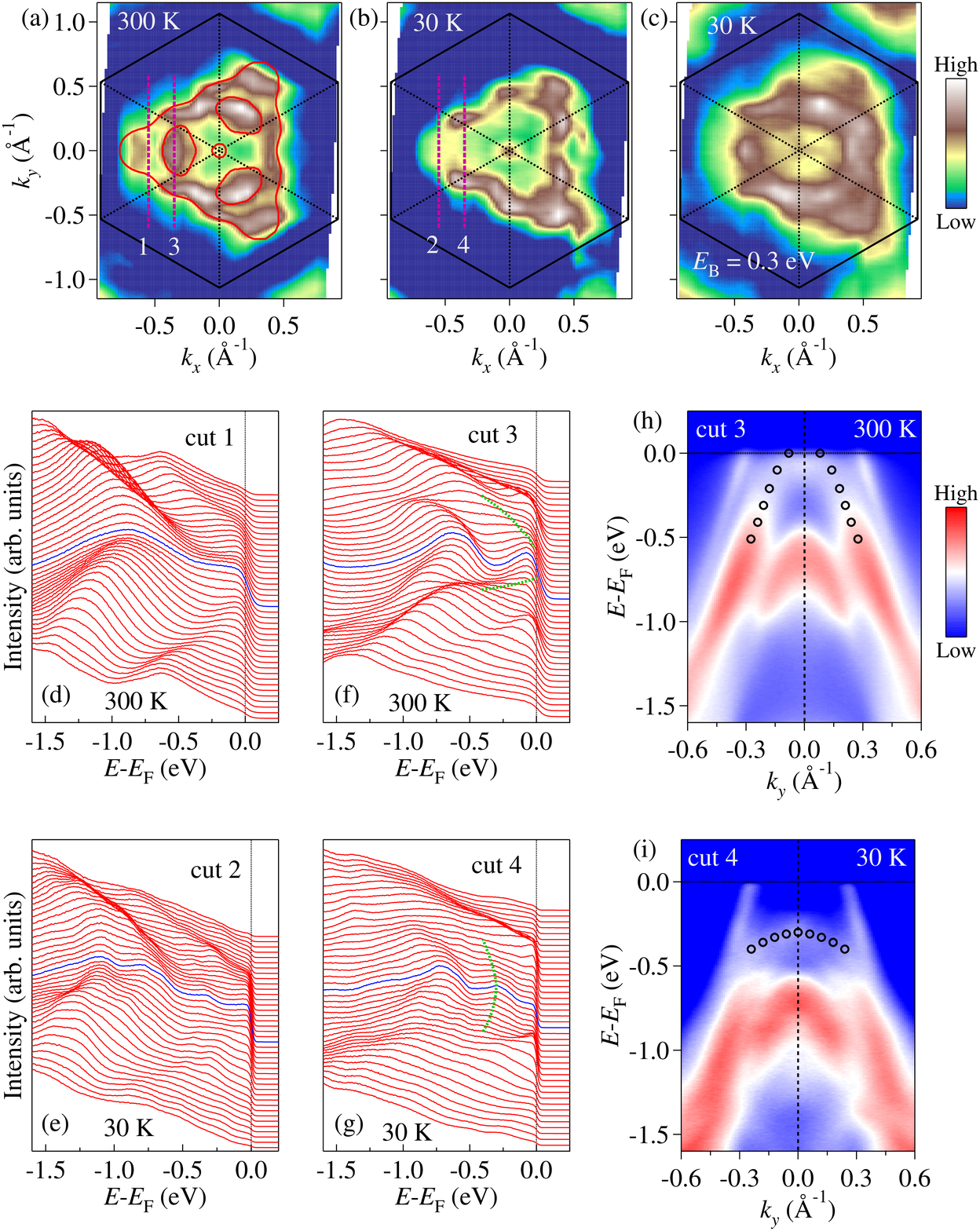}
\caption{(Color online) (a) ARPES intensity plot at \emph{E}$_{F}$ taken with \emph{h}$\nu$ = 90 eV at 300 K, corresponding to \emph{k}$_{z}$ $\sim$ 0.8$\pi$. The intensity is obtained by integrating the spectra within $\pm$10 meV with respect to \emph{E}$_{F}$. The calculated FSs at \emph{k}$_{z}$ = 0.8$\pi$ for the trigonal phase are plotted for comparison. (b) Same as (a) but taken at 30 K. (c) Same as (b) but integrated with respect to \emph{E}$_{B}$ = 0.3 eV. (d) and (e) ARPES spectra recorded along cuts 1 and 2 at \emph{k}$_{x}$ = -0.55 {\AA}$^{-1}$ from panels (a) and (b), respectively. (f) and (g) ARPES spectra recorded along cuts 3 and 4 at \emph{k}$_{x}$ = -0.35 {\AA}$^{-1}$ from panels (a) and (b), respectively. Green dashed lines are guides for eyes to trace the band dispersions. Blue curves indicate the spectra at \emph{k}$_{y}$ = 0. (h) and (i) ARPES intensity plots along cuts 3 and 4, respectively. Open circles are guides for eyes to trace the band dispersions.}
\end{figure}

Having determined the orbital character at the low energy, we turn to the FS topology and band dispersions. Figures 3(a) and 3(b) show the FS intensity maps with photon energy \emph{h}$\nu$ = 90 eV (corresponding to the \emph{k}$_{z}$ $\sim$ 0.8$\pi$ plane) measured at 300 and 30 K, respectively. We observe one outer and three inner hole-like FS pockets at 300 K, in agreement with the calculated ones at \emph{k}$_{z}$ = 0.8$\pi$ for the trigonal phase. According to their calculations, Yang \emph{et} \emph{al}. proposed a FS nesting between one of the corners of the outer pocket at \emph{k}$_{z}$ = 0.8$\pi$ and the inner pocket at \emph{k}$_{z}$ = 0.4$\pi$ with a wavevector of (1/5, 0, -1/5) \cite{ED1}. To inspect whether there exists any energy gap induced by this nesting, we show in Figs. 3(d) and 3(e) the ARPES spectra along the \emph{k}$_{x}$ = -0.55 {\AA}$^{-1}$ cut at 300 and 30 K, respectively. The spectra exhibit a clear Fermi edge at 30 K, suggesting that no CDW-type gap is induced in the proposed nesting FS section, which is consistent with the optical data \cite{optical}. 

The most prominent change in the low temperature phase is that the three inner pockets disappear at 30 K. To clarify this, we compare in Figs. 3(f)-3(i) the band dispersions along the \emph{k}$_{x}$ = -0.35 {\AA}$^{-1}$ cut at 300 and 30 K. While the inner hole-like band crosses \emph{E}$_{F}$ at 300 K, forming the inner FS pocket, it sinks below \emph{E}$_{F}$ at 30 K, leading to a large energy gap of $\sim$ 0.3 eV. From the intensity map at \emph{E}$_{B}$ = 0.3 eV and 30 K in Fig. 3(c), we observe again three highlighted areas, corresponding to the band tops, whose momentum locations are consistent with those of the inner FS pockets at 300 K. This indicates that the change in the FS topology arises from an energy shift of the related band. According to the band calculations, the near-\emph{E}$_{F}$ band that forms the inner pockets originates mainly from the Te \emph{p}$_{x}$+\emph{p}$_{y}$ orbitals. As seen in Fig. 4(b), the partial DOS from the Te \emph{p}$_{x}$+\emph{p}$_{y}$ orbitals exhibits a distinct peak located very close to \emph{E}$_{F}$. The large energy shift of the Te \emph{p}$_{x}$+\emph{p}$_{y}$ band reduces remarkably the DOS at \emph{E}$_{F}$ and the kinetic energy of the electrons, which is likely to be the driving force for the transition. 

\begin{figure}
\includegraphics[clip,width=3in]{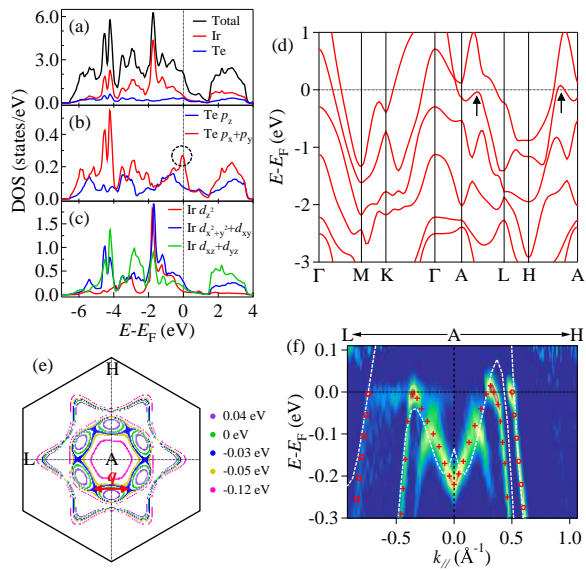}
\caption{(Color online) (a) Total and partial DOS from the Te 5\emph{p} and Ir 5\emph{d} states. (b) Partial DOS from different Te 5\emph{p} orbitals. (c) Partial DOS from different Ir 5\emph{d} orbitals. (d) Calculated band structure for the trigonal phase. The arrows indicate the tops of the band from the Te \emph{p}$_{x}$+\emph{p}$_{y}$ orbitals along AL and AH. (e) Contour map of \emph{E}$_{B}$'s of the bands at \emph{k}$_{z}$ = $\pi$, which is conveniently obtained through the maximally localized Wannier functions derived from the \emph{ab} \emph{initio} calculations. The arrow indicates the momentum transfer connecting the adjacent saddle points. (f) Intensity plot of the two-dimension curvature \cite{curvature} along L-A-H taken with \emph{h}$\nu$ = 96 eV at 300 K. Red symbols are guides for eyes to trace the band dispersions. White dashed curves represent the calculated bands.}
\end{figure}

We further find that the distinct peak in the DOS is a van Hove singularity. According to the calculated band structure in Fig. 4(d), as moving from the A point to the L and H points, the Te \emph{p}$_{x}$+\emph{p}$_{y}$ band disperses first towards \emph{E}$_{F}$ and then to higher \emph{E}$_{B}$, leading to band tops at k$_{//}$ $\sim$ 0.36 {\AA}$^{-1}$ indicated by arrows. This band crosses and sinks below \emph{E}$_{F}$ along AH and AL, respectively, forming six small hole pockets at \emph{k}$_{z}$ = $\pi$. From the contour map of \emph{E}$_{B}$'s of the bands at \emph{k}$_{z}$ = $\pi$ in Fig. 4(e), one can identify six saddle points located along the AL lines, at which the band dispersions are hole- and electron-like along and perpendicular to AL, respectively. It is worth noting that the momentum transfer connecting the adjacent saddle points is \emph{q} $\sim$ (0.19, 0, 0), which is very close to the in-plane projection of the structural modulation wavevector \emph{Q} = (1/5, 0, -1/5). In Fig. 4(f), we compare the experimental band dispersions and the calculated ones along L-A-H. The excellent consistency in the experimental and calculated data supports the scenario of the saddle points, whose energy position in experiment is just located at \emph{E}$_{F}$. 

Our results indicate that the structural transition in IrTe$_{2}$ is intimately associated with the vHs at \emph{E}$_{F}$. Firstly, the wavevector between the adjacent saddle points is very close to the in-plane projection of the structural modulation wavevector. Secondly, the bands near the saddle points are strongly reconstructed, which removes the vHs from \emph{E}$_{F}$ and thus reduces the kinetic energy of the electronic system. This is reminiscent of the Rice-Scott ``saddle-point'' mechanism \cite{NbSe2_Shin,Rice_Scott,saddle}. In this model, the susceptibility diverges logarithmically at a wavevector connecting two saddle points, leading to a CDW stability, and the FS is truncated near the saddle points. Although the phase transition in IrTe$_{2}$ can be qualitatively understood in the framework of the ``saddle-point'' mechanism, distinct deviations from the characteristics of a typical CDW transition are observed. Firstly, the model predicts that the conductivity is enhanced due to the removal of the saddle points, where the effect of the scattering rate reduction is stronger than that of the carrier number reduction. However, a steep jump in the resistivity was observed across \emph{T}$_{s}$ upon cooling in IrTe$_{2}$ \cite{ED1,optical,JPSJ_PS,CuxIrTe2,pressure}. Secondly, the bands are strongly modified in a large energy range of at least 3 eV \cite{optical,ARPES}. Finally, the superstructure modulations are highly nonsinusoidal and rather rectangular \cite{ED2}. While there is no theoretical picture to explain all the experimental results, the deviations could be associated with the strong lattice distortions \cite{structure,ED2} since a Kohn anomaly can be induced by the scattering of conduction electrons between the adjacent saddle points. This suggests that the phase transition is not purely electronically driven and a lattice degree of freedom is intricately coupled with a charge degree of freedom. 

In the general vHs scenario, various states, including CDW, spin-density-wave, phase separation, and even superconductivity, may be induced to remove the singularity from \emph{E}$_{F}$ and reduce the kinetic energy of the electronic system. One interesting question is whether superconductivity is also driven by the vHs since the opening of a superconducting gap in the vicinity of the saddle points could remove the vHs from \emph{E}$_{F}$. As the phase diagrams of doping and pressure exhibit a seemingly competitive interplay between superconductivity and other ordered phases \cite{ED1,optical,JPSJ_PS,CuxIrTe2,RhxIrTe2,pressure}, our present study, which reveals the vHs origin of the phase transition in IrTe$_{2}$, will motivate further experimental and theoretical studies on the relationship between superconductivity and adjacent quantum ordered states.

We acknowledge W. Ku, T. Xiang, Z. Y. Lu, G. L. Cao, H. F. Tian, J. Q. Li, X. D. Zhou, and Y. Y. Wang for valuable discussions. This work was supported by grants from CAS (2010Y1JB6), MOST (2010CB923000, 2011CBA001000, 2011CB921701, 2013CB921700 and 2012CB821403), NSFC (11004232, 11050110422, 11204359, 11120101003, 11074291, 11121063, 11234014, and 11274362), and SSSTC (IZLCZ2 138954). This work at Brookhaven was support by the U.S. Department of Energy, Division of Materials Science, under Contract No. DE-AC02-98CH10886. This work is based in part on research conducted at the Synchrotron Radiation Center, which is primarily funded by the University of Wisconsin-Madison with supplemental support from facility users and the University of Wisconsin-Milwaukee. This work is based in part upon research conducted at the Swiss Light Source, Paul Scherrer Institute, Villigen, Switzerland. The Advanced Light Source and the Advanced Photon Source are supported by the U.S. Department of Energy, Office of Science, Office of Basic Energy Sciences, under Contract No. DE-AC02-05CH112 and DE-AC02-06CH11357, respectively.

\bibliographystyle{apsrev4-1}

\end{document}